\documentclass[twoside]{article}
\usepackage{amsmath,amssymb,amsthm}
\usepackage{qic}
\usepackage{braket}
\usepackage{qcircuit}
\usepackage{tikz}
\usetikzlibrary{shapes,arrows}
\usepackage{hyperref}

\newenvironment{sistema}%
  {\left\lbrace\begin{array}{@{}l@{}}}%
  {\end{array}\right.}

\newcommand{\effe}{$F_k$}

\newcommand{\stringhe}[1]{\{0,1\}^{#1}}
\newcommand{\testo}[1]{\text{#1}}
\newcommand{\concat}{| \! |}
\newcommand{\prefisso}[2]{#1_1 \concat \cdots \concat #1_{#2}}

\newcommand{\cbc}[2]{\testo{CBC}^{#1}_{#2}}

\newcommand{\tuple}[2]{#1_1 , \dots , #1_{#2}}
\newcommand{\abovetuple}[2]{#1^1 , \dots , #1^{#2}}
\newcommand{\suma}{$\bigoplus$}
\newcommand{\img}{\mathrm{img}}

\tikzset{%
  block/.style    = {draw, thick, rectangle, minimum height = 2em,
    minimum width = 2em},
}

\theoremstyle{plain}
\newtheorem{teo}{Theorem}
\theoremstyle{plain}
\newtheorem{defi}[teo]{Definition}

\begin{document}
\setlength{\textheight}{8.0truein}    %FOR 2ND PAGE ONWARDS

% \title{}

% \date{\today}

\runninghead{Using Simon's algorithm to attack symmetric-key cryptographic primitives}
            {T. Santoli and C. Schaffner}

\normalsize\textlineskip
\thispagestyle{empty}
%\setcounter{page}{65}

%\copyrightheading{Vol.}{No.}{Year}{Page Nos.}
%\copyrightheading{17}{1\&2}{2017}{0065--0078}

\vspace*{0.88truein}

\alphfootnote

\fpage{65}

\centerline{\bf
%Using Simon's Algorithm to Attack Symmetric-Key Cryptographic Primitives
USING SIMON'S ALGORITHM TO ATTACK SYMMETRIC-KEY}\vspace*{3pt} \centerline{\bf CRYPTOGRAPHIC PRIMITIVES}
\vspace*{0.37truein}
\centerline{\footnotesize
%Thomas Santoli
THOMAS SANTOLI\footnote{\href{mailto:thsantoli1@gmail.com}{thsantoli1@gmail.com}}}
\vspace*{0.015truein}
\centerline{\footnotesize\it Mathematical Institute, University of Oxford; \texttt{}}
\baselineskip=10pt
\centerline{\footnotesize\it Andrew Wiles Building, Radcliffe Observatory Quarter}
\baselineskip=10pt
\centerline{\footnotesize\it Woodstock Road, Oxford, OX2 6GG}
\vspace*{10pt}
\centerline{\footnotesize 
%Christian Schaffner
CHRISTIAN SCHAFFNER\footnote{\href{mailto:c.schaffner@uva.nl}{c.schaffner@uva.nl}}}
\vspace*{0.015truein}
\centerline{\footnotesize\it ILLC, University of Amsterdam}
\vspace*{0.015truein}
\centerline{\footnotesize\it Centrum Wiskunde \& Informatica (CWI)}
\vspace*{0.015truein}
\centerline{\footnotesize\it \href{http://www.qusoft.org/}{QuSoft}}
\baselineskip=10pt
\centerline{\footnotesize\it P.O. Box 94242, 1090 GE Amsterdam, Netherlands}
\vspace*{0.225truein}
%\publisher{June 15, 2016}{December 23, 2016}

% \author{Thomas Santoli}
% \affiliation{Institute for Logic, Language and Computation (ILLC), University of Amsterdam; \texttt{thsantoli1@gmail.com}} 
% \author{Christian Schaffner}
% \affiliation{ILLC, University of Amsterdam\\ Centrum Wiskunde \& Informatica (CWI)\\ \href{http://www.qusoft.org/}{QuSoft}; \texttt{c.schaffner@uva.nl}.}

\vspace*{0.21truein}

%% \abstracts{first paragraph}{second paragraph}{third paragraph}
%% If there is only one paragraph, just keep the second and third empty 
%% like the following one 
\abstracts{
We present new connections between quantum information and the
field of classical cryptography. In particular, we
provide examples where Simon's algorithm can be used to show
  insecurity of commonly used cryptographic symmetric-key primitives. Specifically, these examples consist of a quantum
  distinguisher for the 3-round Feistel network and a forgery
  attack on CBC-MAC which forges a tag for a chosen-prefix message
  querying only other messages (of the same length). We assume that an adversary has
  quantum-oracle access to the respective classical primitives. Similar results have been achieved recently in independent work by Kaplan \emph{et al.}~\cite{C:KLLN16}. Our findings shed new light on the post-quantum security of
  cryptographic schemes and underline that classical security proofs
  of cryptographic constructions need to be revisited in light of
  quantum attackers.
}{}{}

\vspace*{10pt}

\vspace*{3pt}
%\communicate{R~Jozsa~\&~B~Terhal}

\vspace*{1pt}\textlineskip    %) USE THIS MEASUREMENT WHEN THERE IS
   %) A SECTION HEADING
%\vspace*{-0.5pt}
%\noindent
%%%%%%%%%%%%%%%%%%%%%%%%%%%%%%%%
%put the text of the paper here
%%%%%%%%%%%%%%%%%%%%%%%%%%%%%%%%
\section{Introduction}       
\noindent 
The main goal of cryptography is secure communication. While \emph{encryption} ensures  the
secrecy of a message, its integrity is guaranteed by means of \emph{authentication}. In private-key cryptography, it
is assumed that honest players Alice and Bob share
a private key unknown to any attacker. The situation is symmetric
between Alice and Bob, hence this scenario is also referred to as
\emph{symmetric cryptography}. Here, private-key
encryption (e.g.\ the Advanced Encryption Standard AES in a certain encryption mode)
can be used to ensure secrecy of a message, while
Message-Authentication Codes (MACs) allow to verify that a message
originated from the secret-key holder and that the message has
not been altered in transit. Cryptographic research has come up with
satisfactory solutions for these tasks, but even if we assume that these
problems are solved, the problem of pre-establishing secret keys remains.

In groundbreaking work, very elegant solutions to this key-establishing problem have been
suggested by Merkle~\cite{Mer78} and Diffie and Hellman~\cite{DH76}. The key idea is to split the key up into a
public and private part, where only the latter needs to be kept
secret\footnote{In fact, public-key cryptography
was invented already in the early 70s by Ellis, Cocks and Williams at the British intelligence agency GHCQ, but
  these works had been classified until 1997.}. However, compared to private-key cryptography, public-key
cryptoschemes often require some richer
mathematical structure which makes them more vulnerable to attacks. 

Post-quantum cryptography~\cite{BBD09}\footnote{The term is quite well-established by now, but chosen somewhat unfortunately, because the research area is concerned with cryptography which is still secure \emph{at the beginning} and not after the end of the era of large-scale quantum computers.} studies the security of
cryptographic systems against quantum attackers. Most currently used
public-key systems (such as RSA~\cite{RSA78} or
elliptic-curve cryptography) are known to be insecure against attackers with large-scale quantum computers, because their security relies on the hardness of mathematical problems such as factoring large integer numbers or
taking discrete logarithms. For these problems, there is an
exponential gap between the quantum and classical\footnote{Here and
  throughout the article, we will use the term \emph{classical} as
  synonym for \emph{non-quantum}.} running times of the
best-known algorithms: these problems can be solved in polynomial time by Shor's
algorithm~\cite{Sho94} on a large enough quantum computer, whereas we
do not know of any polynomial-time classical algorithms for them.

In light of the fact that quantum algorithms break most of currently used public-key cryptography, it is natural to study the impact of quantum attacks also on
\emph{private-key} (or \emph{symmetric}) cryptosystems. However, not
much is known about the security of cryptographic primitives (such as
hash functions or encryption schemes) under the
assumption that an attack may involve quantum queries to the
primitive.  So, for example, apart from querying the primitive on a
single input, a quantum adversary is allowed to query it on a superposition of
inputs. Such a model is easily justified in case of a hash
function~\cite{BDF+11} which is given by a classical algorithm. That
algorithm can then be run on a quantum superposition of possible
inputs. Other settings of superposition queries have been
considered as well~\cite{BZ13,DFNS14,EC:BonZha13,C:GagHulSch16}. The common
belief so far was that quantum attacks on symmetric primitives are of minor
concern, as they mainly consist of employing Grover's algorithm~\cite{STOC:Grover96} to
generically speed up search (sub-)problems, for instance for finding collisions in hash
functions. Grover search usually allows to get at most a quadratic improvement
in the number of queries required to attack the primitive, see for
instance~\cite{C:KLLN16} for recent cryptanalytic results in this direction.
In these cases, doubling the security parameter (such as
the bit-size of the output of the hash function) would be a simple
solution to overcome quantum attacks.

In this article, we exhibit two examples of commonly used classical
symmetric-key cryptographic primitives which have well-established
classical security proofs showing that an exponential number of
queries is required for an attack. However, we demonstrate that security is broken by a quantum attacker which only makes a \emph{polynomial}
number of quantum queries. This is a similar \emph{exponential gap} between
classical and quantum security as in the devastating attacks
using Shor's algorithm in the public-key setting. Our results shed new light on the
relation between quantum information and classical cryptography,
suggesting that many classical cryptographic security proofs have been carried out
in an incomplete (namely classical) security model and need to be revisited in
the light of the currently most accurate description of Nature, which
is quantum mechanics.

\subsection{Simon's Algorithm}
Simon's algorithm was proposed in 1994~\cite{FOCS:Simon94} as
first quantum algorithm exhibiting an exponential speedup in query
complexity compared to any classical algorithm (in the bounded-error setting). It solves the
following problem: given a function $f:\{0,1\}^n \rightarrow
\{0,1\}^n$ with the promise that for some $s \in \{0,1\}^n$, it holds
that $[f(x)=f(y)] \Leftrightarrow [x \oplus y \in \{0^n,s\}]$, find
the $n$-bit string $s$. It can be shown~\cite{MdW16} that any
classical algorithm requires $\Omega(2^{n/2})$ classical queries to
$f$ in order to find $s$, whereas Simon's algorithm succeeds using only
$O(n)$ quantum queries to $f$. The quantum part of  Simon's algorithm consists of executing the following circuit:
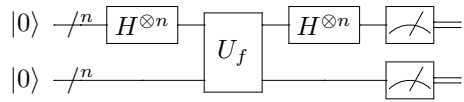
\begin{equation*}
 \Qcircuit @C=1em @R=.7em {
  \lstick{\ket{0}} & /^n \qw & \gate{H^{\otimes n}} & \multigate{1}{U_f} & \gate{H^{\otimes n}} & \meter & \cw \\
  \lstick{\ket{0}} & /^n \qw & \qw                  & \ghost{U_f}        & \qw                  & \meter & \cw }
\end{equation*}
where the quantum query to the classical function $f$ is formalized in the standard way by a unitary transform 
\begin{equation} \label{eq:unitary}
U_f \ket{x}\ket{y} = \ket{x} \ket{y \oplus f(x)}
\end{equation}
where $\oplus$ denotes bitwise addition modulo 2.

The first Hadamard operation leads to the state $\sum_{x \in \{0,1\}^n} \ket{x} \ket{0}$, which after querying $f$ becomes $\sum_{x \in \{0,1\}^n} \ket{x} \ket{f(x)}$. Measuring the second register gives a random outcome $f(x)$ and collapses the first register to $\frac{1}{\sqrt{2}}(\ket{x} + \ket{x \oplus s})$ because of the promised structure of $f$. Applying another Hadamard transform leads to the state $\sum_{y \in \{0,1\}^n} (-1)^{y \cdot x} (1+(-1)^{y \cdot s})\ket{y}$. Note that for $y \in \{0,1\}^n$ with $y \cdot s = y_1 s_1 \oplus \cdots \oplus y_n s_n = 1$, we have destructive interference and the amplitudes of those strings $y$ vanish. Therefore, measuring the register will result in a (random) $n$-bit string $y$ with the property that $y \cdot s = 0$. Running this quantum procedure $O(n)$ of times results in a full-rank system of linear equations which can be solved efficiently classically to obtain the string $s$.

\subsection{Our Contributions} 
In this article, we observe that requiring only the forward implication of Simon's promise $x = y \oplus s \Rightarrow f(x)=f(y)$ suffices for Simon's subroutine to give a $j$ such that $j \cdot s = 0$. The difference between this weakened condition and full Simon's promise is that we cannot bound the number of runs of the subroutine which are required in order to get an independent system of $n$ such equations, and there are cases in which such an independent system cannot be obtained at all. 

Our contributions build on a technique developed by Kuwakado and Morii. They make clever use of Simon's algorithm to distinguish a three-round Feistel construction with internal permutations from a random permutation (see explanations below)~\cite{kuwakado2010quantum}, and to recover the key in the Evan-Mansour encryption scheme~\cite{kuwakado2012security}.
In Section~\ref{sec:feistel}, we slightly extend the result from~\cite{kuwakado2010quantum}. Using our observation above, we can relax the assumption from~\cite{kuwakado2010quantum} that the internal functions have to be permutations, replacing them with arbitrary functions. Using random functions (instead of permutations) is relevant in the application where a Feistel network is used to convert pseudorandom functions into pseudorandom permutations. In Section~\ref{sec:cbc} we present the main new result of this article, a quantum forgery attack on a commonly used scheme for message authentication called CBC-MAC (of fixed-size messages). In our attack, we assume an adversary to be able to query arbitrary quantum superpositions of messages, provided these do not contain the message for which she forges a tag in the end. In fact, our attack also applies to an extension of the basic CBC-MAC to arbitrary-length messages by prepending the length of the message.

\subsection{Related work}
Concurrent to this work, very similar results to ours are presented in a recent article by Kaplan, Leurent, Leverrier, and Naya-Plasencia~\cite{C:KLLN16}. While there has been an initial exchange of ideas in early stages of the work, the communication stopped after a while and our respective results have been obtained independently. 

One difference between our work and \cite{C:KLLN16} is that we use different ways to handle the case in which Simon's promise is not fully satisfied, that is when there is an $s$ such that $f(x) = f(x \oplus s)$ for all inputs $s$, but there can be unwanted collisions $f(x) = f(y)$ where $y \not\in  \{x, x \oplus s\}$. These unwanted collisions may increase the number of runs of Simon's subroutine needed to obtain an independent system of equations. We observe that this does not constitute a problem when it comes to distinguishing $f$ from a random function. Instead, the authors of \cite{C:KLLN16} give a more general analysis of Simon's algorithm, and introduce a quantity to measure the number of the aforementioned unwanted collisions, and they show how to bound the number of needed repetitions of the subroutine in terms of this quantity. 

Another difference between our works is the variant of CBC-MAC which we analyze. In our attack, we focus on MACs for fixed-size messages. In particular, we manage to forge a tag for a message (with a chosen prefix) by querying the MAC with messages of the same length. In \cite{C:KLLN16}, they consider a slight variant of the basic CBC-MAC which can handle arbitrary-size messages by using two separate keys. While we show how to forge a tag for a message in which each block is non-zero by querying only messages which have at least one block of zeros, what they do is showing how to forge a tag for a message which is strictly longer than all messages queried in their attack.

Moreover, the results from \cite{C:KLLN16} illustrate that Simon's algorithm can be used to break not only CBC-MAC, but a whole range of modes of operation for authentication and authenticated encryption. Furthermore, Simon's algorithm can be used to speed up \emph{slide attacks}, a cryptanalytic technique against classical symmetric cryptoschemes.

\section{Pseudorandom Functions and Permutations} \label{sec:prf}
In this section, we provide formal definitions and statements about classical cryptography, following the lines of~\cite{KL07}.

\newcommand{\allstrings}{\{ 0,1 \}^{*}}
\newcommand{\functions}{\mathtt{Func}_n}
\newcommand{\permutations}{\mathtt{Perm}_n}
\newcommand{\oracle}[1]{\mathcal{O}(#1)}
\newcommand{\prob}[2]{\underset{#1 \leftarrow #2}{\mathtt{Pr}}}
\newcommand{\neglig}{\mathtt{negl}(n)}

In the following, we consider \emph{keyed functions}. These are two-input functions $F: \allstrings \times \allstrings \to \allstrings$. Given a keyed function $F$, we say that it is \emph{efficient} if there exists a polynomial-time algorithm which computes $F(k,x)$ given $k$ and $x$ as inputs, and we say that it is \emph{length-preserving} if it accepts only pairs of inputs $(k,x)$ where $k$ and $x$ have the same length, and if also the output $F(k,x)$ has this same length.

For each $n \in \mathbb{N}$, let $\functions$ be the set of all functions mapping $n$-bit strings to $n$-bit strings. A length-preserving keyed function $F$ induces a distribution on $\functions$, given by randomly choosing $k \leftarrow \{ 0,1 \}^n$ and then considering the single-input function $F_k(\cdot):=F(k,\cdot)$.

Intuitively we say that $F$ is pseudorandom if, choosing $k$ at random in $\{ 0,1 \}^n$, the function $F_k$ cannot be efficiently distinguished from a function $f$ chosen uniformly at random from the set of all functions $\functions$. In order to give the formal definition of pseudorandom function, we use the notion of \emph{oracle-distinguisher}, which consists of an algorithm $D^{\oracle{\cdot}}$ having access to an oracle function $\oracle{\cdot}$. This means that in one time step, $D$ can query the oracle at a point $x$, obtaining the value of $\oracle{x}$.

\begin{defi}
  Let $F : \allstrings \times \allstrings \to \allstrings$ be an efficient length-preserving keyed function. We call $F$ a \emph{pseudorandom function} if for any probabilistic polynomial-time oracle-distinguisher $D$ there exists a negligible function $\neglig$ such that
  \begin{align*}
  \Big| \prob{k}{\{ 0,1 \}^n} [ D^{F_k(\cdot)}&(1^n) = 1 ]  -   \prob{f}{\functions}[ D^{f(\cdot)}(1^n) = 1 ]   \Big| \leq \neglig 
  \end{align*}
  where a negligible function is one which grows slower than any inverse polynomial in $n$.
\end{defi}

An efficient length-preserving function $F : \allstrings \times \allstrings \to \allstrings$ is called \emph{keyed permutation} if for every $n$ and for every $k \in \{ 0,1 \}^n$ the function $F_k(\cdot) : \{ 0,1 \}^n \to \{ 0,1 \}^n$ is a permutation (that is, a bijection). A keyed permutation is called \emph{efficient} if there are polynomial-time algorithms which, on inputs $k$ and $x$, compute $F_k(x)$ and $F_k^{-1}(x)$ respectively.

Let $\permutations$ be the set of all permutations of $\{ 0,1 \}^n$. In the same way as a length-preserving keyed function induces a distribution on $\functions$, a keyed permutation induces a distribution on $\permutations$. This leads to the definition of pseudorandom permutations, analogous to pseudorandom functions:

\begin{defi}
  Let $F : \allstrings \times \allstrings \to \allstrings$ be an efficient keyed permutation. We call $F$ a \emph{pseudorandom permutation} if for any probabilistic polynomial-time oracle-distinguisher $D$ there exists a negligible function $\neglig$ such that
  \begin{align*}
  \Big| \prob{k}{\{ 0,1 \}^n} [ D^{F_k(\cdot)}&(1^n) = 1 ]
  - \prob{\pi}{\permutations}[ D^{\pi(\cdot)}(1^n) = 1 ]   \Big| \leq \neglig . 
  \end{align*}
\end{defi}

We can use the Feistel construction to build a pseudorandom permutation from a pseudorandom function:
\begin{teo}\cite{LubRac88} \label{thm:feistel}
  If $F$ is a pseudorandom function, then the 3-round Feistel network (as in Figure~\ref{fig:Feistel}) with internal round functions $F_{k_1},F_{k_2},F_{k_3}$ is a pseudorandom permutation mapping $2n$-bit strings to $2n$-bit strings, and using a key $k=k_1 \concat k_2 \concat k_3$ of length $3n$.
\end{teo}

The notion of pseudorandomness of classical functions can be generalized to a post-quantum setting, where we assume that adversaries are allowed to perform quantum computation. In this context, we consider \emph{quantum oracle-distinguishers}. A quantum oracle-distinguisher is a quantum algorithm $D^{\ket{\oracle{\cdot}}}$ which is allowed to make quantum queries of the form $\ket{x,y} \mapsto \ket{x,y \oplus \oracle{x}}$. This means that $D$ can evaluate the function $\oracle{\cdot}$ not only on classical inputs, but also on superpositions of classical inputs. This leads to the following quantum generalizations of pseudorandomness:

\begin{defi} \label{qPRF}
  Let $F : \allstrings \times \allstrings \to \allstrings$ be an efficient length-preserving keyed function. We call $F$ a \emph{quantum pseudorandom function} if for any polynomial-time quantum oracle-distinguisher $D$ there exists a negligible function $\neglig$ such that
  \begin{align*}
  \Big| \prob{k}{\{ 0,1 \}^n} [ D^{\ket{F_k(\cdot)}}&(1^n) = 1 ]  
  - \prob{f}{\functions}[ D^{\ket{f(\cdot)}}(1^n) = 1 ]   \Big| \leq \neglig .
  \end{align*}
  If $F : \allstrings \times \allstrings \to \allstrings$ is an efficient keyed permutation, we call it a \emph{quantum pseudorandom permutation} if for any polynomial-time quantum oracle-distinguisher $D$ there exists a negligible function $\neglig$ such that
  \begin{align*}
  \Big| \prob{k}{\{ 0,1 \}^n} [ D^{\ket{F_k(\cdot)}}&(1^n) = 1 ]  
  - \prob{\pi}{\permutations}[ D^{\ket{\pi(\cdot)}}(1^n) = 1 ]   \Big| \leq \neglig . 
  \end{align*}
\end{defi}

Quantum pseudorandomness is a strictly stronger requirement than classical pseudorandomness. For example, by the distinguisher which we showed in Section~\ref{sec:feistel}, the 3-round Feistel network is not a quantum pseudorandom permutation, while by Theorem~\ref{thm:feistel} it is a classical pseudorandom permutation.

\section{Quantum distinguisher for the 3-round Feistel network} \label{sec:feistel}
A Feistel network is a common cryptographic way of building block ciphers. Block ciphers are the basic building blocks of encryption schemes. Ideally, a block cipher with block size $n$ bits behaves like a fully random permutation from $\{0,1\}^n$ to $\{0,1\}^n$. Many block ciphers such as DES (Data Encryption Standard) are based on a Feistel construction. 

A Feistel network operates in a series of $k$ rounds where internal randomizing functions $\tuple{P}{k}$ (from $n$ bits to $n$ bits) are cleverly put together to result in a permutation. More specifically, if the input of the $i$-th round is $L_{i-1} \concat R_{i-1} \in \{ 0,1 \}^{2n}$, the output produced by the round is $L_i \concat R_i \in \{ 0,1 \}^{2n}$ defined by $L_i := R_{i-1}$ and $R_i:= L_{i-1} \oplus P_i(R_{i-1})$, see Figure~\ref{fig:Feistel}.
Interestingly, this classical cryptographic construction is not alien to quantum information theorists, as it is akin to the standard way of implementing a classical function $P_i$ in a unitary way as in Equation~\eqref{eq:unitary}.

Thus, a $k$-round Feistel network takes an input $L_0 \concat R_0 \in \{ 0,1 \}^{2n}$ and produces an output $L_k \concat R_k \in \{ 0,1 \}^{2n}$ by subsequently executing the round functions. One important feature of Feistel networks is that, regardless of the internal functions $\tuple{P}{k}$, the resulting function $L_0 \concat R_0 \mapsto L_k \concat R_k$ is always a permutation. Indeed, in order to invert the function one only needs to reverse the construction without the need of computing the inverse of the internal functions $\tuple{P}{k}$ (in other words, the internal functions do not have to be bijective).  This is also handy in practice, because it makes the decryption process very similar to the encryption process.

\begin{figure}
\begin{center}
  \begin{tikzpicture}[auto, thick, node distance=2cm, >=triangle 45]
  \draw
  node at (0,0) (l1) {$L_0$}
  node at (4,0) (r1) {$R_0$}
  node at (0,-1) (s1) {\suma} 
  node [block, right of= s1] (p1) {$P_1$}
  node at (0,-3) (s2) {\suma} 
  node [block, right of= s2] (p2) {$P_2$}
  node at (0,-5) (s3) {\suma} 
  node [block, right of= s3] (p3) {$P_3$}
  node at (0,-7) (l3) {$L_3$}
  node at (4,-7) (r3) {$R_3$}
  ;
  \draw[->](l1) -- (s1);
  \draw[->](p1) -- (s1);
  \draw[->](p2) -- (s2);
  \draw[->](p3) -- (s3);
  \draw[->](r1) |- (p1);
  \draw[->](r1) -- (4,-1.5) -- (0,-2) -- node {$L_1$} (s2);
  \draw[->](4,-3) -- (4,-3.5) -- (0,-4) -- node {$L_2$} (s3);
  \draw[->](4,-5) -- (4,-5.5) -- (0,-6) -- (l3);
  \draw[->](s1) -- (0,-1.5) -- (4,-2) -- node {$R_1$} (4,-3) -- (p2);
  \draw[->](s2) -- (0,-3.5) -- (4,-4) -- node {$R_2$} (4,-5) -- (p3);
  \draw[->](s3) -- (0,-5.5) -- (4,-6) -- (r3);
\end{tikzpicture}
\end{center}
\fcaption{A three-round Feistel network with internal functions $P_1$, $P_2$, $P_3$. \label{fig:Feistel}}
\end{figure}

Classically, Theorem~\ref{thm:feistel} states that a 3-round Feistel network constitutes a cryptographically secure pseudorandom permutation as long as the internal functions are pseudorandom as well. That is, a polynomial-time classical distinguisher cannot distinguish a 3-round Feistel network from a random permutation, see Section~\ref{sec:prf} for formal definitions and statements of this classical result. Quantumly, this implication does not hold, and in this section we show a polynomial-time quantum attacker which can distinguish a 3-round Feistel network (with internal random functions) from a fully random permutation. Our proof follows the lines of the results by Kuwakado and Morii~\cite{kuwakado2010quantum}. The main difference is that they assume the internal functions of the Feistel network to be permutations, while here we can handle the more general scenario of arbitrary (random) functions.

In this scenario, a distinguishing adversary has an oracle $V:\stringhe{2n} \to \stringhe{2n}$ which is either a fully random permutation of $2n$-bit strings, or a 3-round Feistel network with internal random functions $P_1, P_2, P_3$, that is 
\[  V(a \concat c ) = c \oplus P_2(a \oplus P_1(c)) \ \concat \ a \oplus P_1(c) \oplus P_3(c \oplus P_2(a \oplus P_1(c))) \, . \]

Let $W:\stringhe{2n} \to \stringhe{n}$ be the first $n$ bits of $V$: 
$W( a \concat c ) = c \oplus P_2(a \oplus P_1(c))$. Let $\alpha,\beta \in \stringhe{n}$ be distinct fixed strings, and let us consider the function $f:\stringhe{n+1} \to \stringhe{n}$ defined as
\[ f( b \concat a) := \begin{sistema}
	W( a \concat \alpha) \oplus \alpha \quad \testo{ if } b=0 \\[1mm]
	W( a \concat \beta) \oplus \beta \quad \testo{ if } b=1
\end{sistema}  \]
for all $b \in \{0,1\}$ and $a \in \stringhe{n}$. In case $V$ is the Feistel network, we have 
\begin{align*} 
f( b \concat a) &= \begin{sistema}
	\alpha \oplus P_2(a \oplus P_1(\alpha)) \oplus \alpha \quad \testo{ if } b=0 \\[1mm]
	\beta \oplus P_2(a \oplus P_1(\beta)) \oplus \beta \quad \testo{ if } b=1
\end{sistema}\\
&= \begin{sistema}
	P_2(a \oplus P_1(\alpha)) \quad \testo{ if } b=0 \\[1mm]
	P_2(a \oplus P_1(\beta)) \quad \testo{ if } b=1
\end{sistema} \quad .
\end{align*}

For each $y$ in the image $\img(P_2)$ of $P_2$, denote its pre-image by $P_2^{-1}(y) = \{ \abovetuple{x}{l_y} \}$, we have
\begin{align*}  
f(b \concat a) &= y \\ 
\Leftrightarrow 
b \concat a &\in \{ 0 \concat x^i \oplus P_1(\alpha) \}_{i = 1 \dots l_y} \cup \{ 1 \concat x^i \oplus P_1(\beta) \}_{i = 1 \dots l_y}\\
&= \{ \abovetuple{a}{l_y} \} \cup \{ a^1 \oplus s , \dots , a^{l_y} \oplus s \}
\end{align*}
where $s = 1 \concat ( P_1(\alpha) \oplus P_1(\beta) )$ and $a^i = 0 \concat (x^i \oplus P_1(\alpha))$ for $i=1 \dots l_y$. In particular, in case $V$ is a Feistel network, for each $u \in \stringhe{n+1}$, we have $f(u) = f(u \oplus s)$. On the other hand, in case $V$ is a random permutation, no such relation is expected to hold.

\vspace{0,5cm}

Using at most two invocations of the unitary operator $U_V : \ket{x,y} \mapsto \ket{x, y \oplus V(x)}$ for computing $V$, we can construct the unitary operator $U_f : \ket{u,y} \mapsto \ket{u, y \oplus f(u) }$. Using this operator, one can mount the following attack, based on Simon's algorithm, to distinguish whether $V$ is a random permutation or a 3-round Feistel network:

\begin{enumerate}
	\item Initialize $X$ as the empty set. 
	\item Run Simon's subroutine:
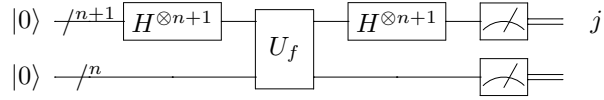
\begin{equation*}
 \Qcircuit @C=1.3em @R=.7em {
&&&  \lstick{\ket{0}} & /^{n+1} \qw & \gate{H^{\otimes n+1}} & \multigate{1}{U_f} & \gate{H^{\otimes n+1}} & \meter & \cw & j \\
&&& \lstick{\ket{0}} & /^n \qw & \qw                  & \ghost{U_f}        & \qw                  & \meter & \cw }
\end{equation*}
	\item Add the result $j = (j_0, \tuple{j}{n}) $ to $X$. If $X$ does not contain $n$ linearly independent $j$'s, go back to Step 2, unless $X$ already contains $2n$ elements. In the latter case, guess $V$ to be a 3-round Feistel network. Otherwise, solve this linear system of equations:
	\[ \begin{pmatrix}
		\ & | & \ \\
		j_1 & \dots & j_n \\
		\ & | & \
	\end{pmatrix} \begin{pmatrix}
		s_1 \\
		\vdots \\
		s_n
	\end{pmatrix} = \begin{pmatrix}
		| \\
		j_0 \\
		|
	\end{pmatrix} \]
	\item Choose $u \in \stringhe{n+1}$ at random, and guess $V$ to be a 3-round Feistel network if $f(u) = f(u \oplus s)$, where $s = (1, \tuple{s}{n})$. Otherwise guess $V$ to be a random permutation.
\end{enumerate} 

Let us first analyze the case when $V$ is a 3-round Feistel network. After applying $U_f$, we have the state
\begin{align*} 
\frac{1}{\sqrt{2^{n+1}}} &\sum_{u \in \stringhe{n+1}} \ket{u,f(u)} = \frac{1}{\sqrt{2^{n+1}}} \sum_{y \in \img(P_2)} \sum_{i=1}^{l_y} ( \ket{a^i} + \ket{a^i \oplus s} ) \ket{y}  \ . 
\end{align*}

Before measuring, we get 
\begin{align*}
&\frac{1}{2^{n+1}} \sum_{y \in \img(P_2)} \sum_{i=1}^{l_y} \Big( \sum_{j \in \stringhe{n+1}} (-1)^{a^i \cdot j} \ket{j} + \sum_{j \in \stringhe{n+1}} (-1)^{(a^i \oplus s)\cdot j} \ket{j}      \Big) \ket{y} \\
=  &\frac{1}{2^{n+1}} \sum_{y \in \img(P_2)}  \sum_{j \in \stringhe{n+1}} \sum_{i=1}^{l_y}  (-1)^{a^i \cdot j} ( 1 + (-1)^{s \cdot j}  ) \ket{j} \ket{y} \\
= &\frac{1}{2^n} \sum_{y \in \img(P_2)} \Big( \sum_{\substack{j \in \stringhe{n+1} \\ s \cdot j = 0}} \Big( \sum_{i=1}^{l_y} (-1)^{a^i \cdot j} \Big) \ket{j} \Big) \ket{y} . 
\end{align*}

Hence, all measurement outcomes $j$ satisfy $s \cdot j = 0 \mod 2$, so if at some point we solve the system in Step~3 we get $(\tuple{s}{n}) = P_1(\alpha) \oplus P_1(\beta)$, and the condition $f(u) = f(u \oplus s)$ in Step~4 is satisfied for all $u \in \stringhe{n+1}$, so the algorithm gives the correct answer. If we do not reach the step of solving the system because after collecting $2n$ vectors we still do not have $n$ independent vectors, then we give the correct answer because we will guess $V$ to be the 3-round Feistel network. \\

If $V$ is a random permutation, then the $j$'s will be uniform random vectors. In that case, we make an error if we do not have $n$ independent vectors after $2n$ extractions, or if $f(u) = f(u \oplus s)$ in Step~4. Notice that the probability that after $2n$ extractions of uniform vectors, we still do not have $n$ independent vectors is negligible in $n$.\footnote{This can be proven as follows: the probability that $n$ uniform vectors of length $2n$ are linearly independent is exactly $\Pi_{i=1}^n(1-2^{i-1-2n}) \geq 1-2^{-n}$. In other words, the column rank of a random binary $2n \times n$ matrix is $n$ except with probability at most $2^{-n}$. The claim then follows directly from the equality of row and column rank, see e.g.~\cite{War05}.} In case we continue to solve a random system of independent equations in Step~3, we get a random vector $\tuple{s}{n}$ as a result. Then, at Step~4, the algorithm makes an error if the randomly chosen $u \in \{ 0,1 \}^{n+1}$ satisfies the condition $f(u)=f(u \oplus s)$, but this also happens only with negligible probability. So we conclude that also if $V$ is a random permutation, the algorithm gives a wrong answer by guessing it to be a 3-round Feistel network with only negligible probability.

In summary, we have constructed a quantum adversary which can efficiently distinguish a 3-round Feistel construction with internal functions from a fully random permutation. This extends the results from~\cite{kuwakado2010quantum} to the important setting where a Feistel construction is used to turn pseudorandom functions into pseudorandom permutations.

%%%%%%%%%%%%%%%%%%%%%%%%%%%%%%%%%%%%%%%%%%%%%%%%%%%%%%%%%%%%%%%%%%%%%%

\section{Quantum forgery attack on CBC-MAC} \label{sec:cbc}

The goal of message authentication and integrity is to allow
communicating parties to verify that a received message comes from the
expected sender, and that the message has not been altered in transit. Parties who share a secret key can use Message Authentication Codes (MACs) to achieve this goal: the sender feeds the key $k$ and the message $m$ to an algorithm $\mathsf{Mac}(k,m)$ which computes a tag $t$ for the message, and sends the pair $(m,t)$ to the receiver. Then the receiver, using the same key $k$, can check whether $t$ is a valid tag for $m$. A message authentication code is considered secure if no efficient adversary who can ask arbitrary adaptive queries to the oracle $\mathsf{Mac}(k,\cdot)$ (where $k$ is uniform and unknown to the adversary) can produce a valid tag for a message $m$ that has not been queried before.

\medskip

CBC-MAC is an efficient way of constructing a message authentication code based on a pseudorandom permutation (formally defined in Section~\ref{sec:prf}). CBC-MAC is widely used in practice, and comes in many variants. Given a pseudorandom permutation $F: K \times \stringhe{n} \to \stringhe{n}$, the basic CBC-MAC construction is defined as
\begin{align*}  
\cbc{\ell}{ } : K \times \stringhe{\ell n} & \to \stringhe{n}  \\
(k,m_1 \concat m_2 \concat \dots \concat m_\ell) & \mapsto \cbc{\ell}{F_k}(m_1 \concat m_2 \concat \dots \concat m_\ell)  \, ,
\end{align*}
where
\begin{align*}  
&\cbc{\ell}{F_k}(m_1 \concat m_2 \concat \dots \concat m_\ell) = F_k(F_k( \dots F_k(F_k(m_1) \oplus m_2 ) \oplus \dots ) \oplus m_\ell  ) \, .
\end{align*}
The case $\ell=3$ is depicted in Figure~\ref{fig:cbc}.

\medskip

\begin{figure}
\begin{center}

\begin{tikzpicture}[auto, thick, node distance=2cm, >=triangle 45]
  \draw
  node at (0,-2) [block, name=effe1] {\effe}
  node [block, right of= effe1] (effe2) {\effe}
  node [block, right of= effe2] (effe3) {\effe}
  node at (2,-1) (somma1) {\suma}
  node [right of= somma1] (somma2) {\suma}
  node at (0,0) (input1) {$m_1$}
  node [right of= input1] (input2) {$m_2$}
  node [right of= input2] (input3) {$m_3$}
  node at (4,-3) [name=output] {$t$}
  ;
  \draw[->](input1) --  (effe1);
  \draw[->](input2) --  (somma1);
  \draw[->](input3) --  (somma2);
  \draw[->](somma1) --  (effe2);
  \draw[->](somma2) --  (effe3);
  \draw[->](effe1) |- (1,-3) |- (somma1);
  \draw[->](effe2) |- (3,-3) |- (somma2);
  \draw[->](effe3) -- (output);
\end{tikzpicture}

\end{center}
\fcaption{Illustration of the computation of $t = CBC^3_{F_k}(m_1 \concat m_2 \concat m_3)$. \label{fig:cbc}}
\end{figure}

\medskip

Using this construction, a sender who wants to send a message $m$ of length $\ell n$ can compute a tag $t:=\cbc{\ell}{ } (k,m)$ and send the tuple $(m,t)$ to the receiver who can verify whether $t=\cbc{\ell}{ } (k,m)$.

\medskip

One can prove that $\cbc{\ell}{ }$ is a pseudorandom function. Therefore, as explained in~\cite{BKR00}, it (classically) constitutes a secure MAC. Quantumly, even if $F$ is a quantum-secure pseudorandom permutation, this construction does not yield a secure MAC, and not even a quantum pseudorandom function (qPRF) as defined in Section~\ref{sec:prf}.

According to the definition given in~\cite{EC:BonZha13}, a signature scheme is quantum secure if there is no quantum attacker who can produce $q+1$ valid tags by making only $q$ queries. This notion of security is also the one used in~\cite{C:KLLN16}. In this work, we use a different notion of security. Namely, we show how to forge a valid tag for a chosen-prefix message without ever querying that particular message (not even as small part of a big superposition query).
%
 % In this work, we exhibit a different kind of quantum attack which produces a valid tag for a message $m$ with chosen prefix without ever querying that $m$ to the signature oracle (not even as small part of a big superposition query).
 % instead, with a different notion of security. Namely, we consider a signature scheme quantum secure if no quantum attacker can produce a valid tag for a message $m$ without quering the particular message $m$.
%
We note that basic CBC-MAC is only secure for fixed-length messages, but fails in case the attacker is allowed to query and forge different-length messages. However, we would like to stress that our quantum attack does not rely on that weakness, but also breaks the extension of CBC-MAC where the length of the messages is prepended. This extension can be proven secure classically.

In the following, we describe our forgery attack on CBC-MAC for messages of fixed length $\ell n$, where $\ell \geq 3$. We want to forge a tag for a message which has some fixed prefix $\beta \in \{0,1\}^{kn}$ of length $kn$ for some $k \leq \ell-2$. For simplicity, we restrict ourselves to the case $k=1$. In Section~\ref{subsec:extension}, we explain how to extend the attack to prefixes of arbitrary size $k \leq \ell-2$.

In a first step, we assume that the pseudorandom permutation $F_k$ is replaced with a fully random permutation $\pi$. If this replacement made a difference in our forging algorithm, then this difference would allow us to use our algorithm to distinguish $F_k$ from a fully random permutation, in contradiction with the fact that $F$ is a qPRF. 

\medskip

Suppose we want to tag a message having $\alpha_1 \in \stringhe{n}$ as a prefix. Choose some $\alpha_0 \neq \alpha_1$.
For $j = 1 \dots \ell-1$, let us consider the following functions: 

\begin{align*}
  g_j : &\stringhe{n+1} \to \stringhe{n} \\
  &b \concat x \mapsto \cbc{\ell}{\pi} (\alpha_b \concat 0^{(j-1)n} \concat x \concat 0^{(\ell-j  -1  )n}) = \pi^{\ell-j}(\pi^j(\alpha_b) \oplus x) \\
\end{align*}

Note that, for distinct $u,v \in \stringhe{n+1}$, we have $g_j(u) = g_j(v)$ if and only if $v= u \, \oplus \,  1 \concat \!\left(\pi^j(\alpha_0) \oplus \pi^j(\alpha_1)\right)$.

Using one invocation of the unitary operator $U_{\cbc{\ell}{\pi}}$ we can construct the unitary operator $U_{g_j}$, for $j=1 \dots \ell-1$. A quantum adversary can use these operators to forge a tag on a message without querying it as follows: 
\begin{enumerate}
  \item For $j=1 \dots \ell-1$, run Simon's subroutine on $U_{g_j}$ in order to find $z^j := \pi^j(\alpha_0) \oplus \pi^j(\alpha_1)$.\footnote{This step of the algorithm shows that also for the case $\ell=2$, $\cbc{\ell}{\pi}$ is not a quantum pseudorandom permutation: We can run Simon's algorithm with $g_1$ (which exists also in the case $\ell=2$) to get a solution $1 \concat z^1$. Then, we pick $u$ at random and check if  $g_1(u)=g_1(u \oplus 1 \concat z^1)$ holds. For $\cbc{\ell}{\pi}$, this will always be the case, whereas for a fully random permutation, the check will only be passed with negligible probability.}
  Note that all queries made in this process are on (superpositions of) messages $m_1 \concat m_2 \concat \dots \concat m_\ell$ such that $m_i=0^n$ for at least one $i \in \{ 1 \dots \ell \}$ \footnote{This would not be the case for $\ell=2$, which is why we require $\ell \geq 3$ for the attack.}.
  \item Classically query $t_0 := \cbc{\ell}{\pi}(\alpha_0 \concat  0^{(\ell-1)n}) =  \pi^\ell(\alpha_0)$ and $t_1 := \cbc{\ell}{\pi}(\alpha_1 \concat  0^{(\ell-1)n}) = \pi^\ell(\alpha_1)$. \\
  Also these queries are made on messages $m_1 \concat m_2 \concat \dots \concat m_\ell$ such that $m_i=0^n$ for at least one $i \in \{ 1 \dots \ell \}$.
  \item If $\ell$ is even, forge $(m,t) := (\alpha_1 \concat z^1 \concat z^2 \concat \dots \concat z^{\ell-1}, t_0)$. If $\ell$ is odd, forge $(m,t) := (\alpha_1 \concat z^1 \concat z^2 \concat \dots \concat z^{\ell-1}, t_1)$.\\
\end{enumerate}

The message $m=\alpha_1 \concat z^1 \concat z^2 \concat \dots \concat z^{\ell-1}$ is never queried in the above procedure, where we query only messages $m_1 \concat m_2 \concat \dots \concat m_\ell$ such that $m_i=0^n$ for at least one $i \in \{ 1 \dots \ell \}$. In fact, as the $\pi^i$ are permutations, we have $z^j = \pi^j(\alpha_0) \oplus \pi^j(\alpha_1) \neq 0^n$ for all $j$'s.

Therefore, in order to show that the above forging attack is successful, it remains to prove that $t$ is a valid tag for $m$, which means that $\cbc{\ell}{\pi}(m) = t$, where $t=t_0 = \pi^\ell(\alpha_0)$ if $\ell$ is even and $t=t_1=\pi^\ell(\alpha_1)$ if $\ell$ is odd. This claim is proven by induction on $\ell \geq 3$: 

We have two base cases, $\ell = 3$ and $\ell = 4$.  For $\ell=3$ we have:
\begin{align*}  
&\cbc{3}{\pi}(\alpha_1 \concat z^1 \concat z^2) = \pi(\pi(\pi(\alpha_1) \oplus z^1) \oplus z^2) \\
&= \pi\left(\pi\left(\pi(\alpha_1) \oplus \pi(\alpha_0) \oplus \pi(\alpha_1) \right) \oplus \pi^2(\alpha_0) \oplus \pi^2(\alpha_1) \right) \\
&= \pi( ( \pi^2( \alpha_0 ) \oplus  \pi^2(\alpha_0) \oplus \pi^2(\alpha_1) ) = \pi^3(\alpha_1) = t_1 \ .
\end{align*}
For $\ell =4$ we similarly argue that
\begin{align*}
&\cbc{4}{\pi}(\alpha_1 \concat z^1 \concat z^2 \concat z^3) = \pi(\pi(\pi(\pi(\alpha_1) \oplus z^1) \oplus z^2) \oplus z^3)\\
&=\pi\Big(\pi(\pi(\pi(\alpha_1) \oplus \pi(\alpha_0) \oplus \pi(\alpha_1) ) \oplus \pi^2(\alpha_0) \oplus \pi^2(\alpha_1))
\oplus \pi^3(\alpha_0) \oplus \pi^3(\alpha_1) \Big) \\
&= \pi(\pi(  \pi^2( \alpha_0 ) \oplus  \pi^2(\alpha_0) \oplus \pi^2(\alpha_1) ) \oplus \pi^3(\alpha_0) \oplus \pi^3(\alpha_1) )\\
&= \pi( \pi^3( \alpha_1 ) \oplus  \pi^3(\alpha_0) \oplus \pi^3(\alpha_1) ) = \pi^4(\alpha_0) = t_0 \ .
\end{align*}

If $\ell+1$ is even then $\ell$ is odd, and the induction step works as 
\begin{align*}
&\cbc{\ell+1}{\pi} (\alpha_1 \concat z^1 \concat z^2 \concat \dots \concat z^{\ell-1} \concat z^\ell)\\
&= \pi( \cbc{\ell}{\pi}(\alpha_1 \concat z^1 \concat z^2 \concat \dots \concat z^{\ell-1}) \oplus z^\ell )\\
&= \pi( \pi^\ell(\alpha_1) \oplus \pi^\ell(\alpha_0) \oplus \pi^\ell(\alpha_1) ) = \pi^{\ell+1}(\alpha_0) = t_0 \ .
\end{align*}
Otherwise, if $\ell+1$ is odd, then $\ell$ is even and one can argue similarly that
\begin{align*}
&\cbc{\ell+1}{\pi} (\alpha_1 \concat z^1 \concat z^2 \concat \dots \concat z^{\ell-1} \concat z^\ell)\\
&= \pi( \cbc{\ell}{\pi}(\alpha_1 \concat z^1 \concat z^2 \concat \dots \concat z^{\ell-1}) \oplus z^\ell )\\
&= \pi( \pi^\ell(\alpha_0) \oplus \pi^\ell(\alpha_0) \oplus \pi^\ell(\alpha_1) ) = \pi^{\ell+1}(\alpha_1) = t_1 \ .
\end{align*}

Thus, we can conclude by induction that the output pair $(m,t)$ is correct.

\subsection{Extension of the Attack on CBC-MAC} \label{subsec:extension}
The algorithm presented in this section can be adapted for forging a tag for a message having arbitrary $\prefisso{\beta}{k} \in \stringhe{kn}$ as a prefix, for some $k \leq \ell-2$.

Choose $\prefisso{\alpha}{k}$ such that 
\begin{align*}
\overline{\alpha} &:=  \pi(\cdots \pi( \pi(\alpha_1) \oplus \alpha_2 ) \cdots) \oplus \alpha_k 
\neq \pi(\cdots \pi( \pi(\beta_1) \oplus \beta_2 ) \cdots) \oplus \beta_k  =: \overline{\beta} \, .
\end{align*}
This inequality can be achieved by repeatedly choosing $\prefisso{\alpha}{k}$ at random and querying $\cbc{\ell}{\pi} (\prefisso{\alpha}{k} \concat 0^{(\ell-k)n} ) = \pi^{\ell-k} \Big( \pi( \pi(\cdots \pi( \pi(\alpha_1) \oplus \alpha_2 ) \cdots) \oplus \alpha_k ) \Big) = \pi^{\ell-k}(  \pi (  \overline{\alpha}  )  )  = \pi^{\ell-k+1}(\overline{\alpha})  $ and $\cbc{\ell}{\pi}(\prefisso{\beta}{k} \concat 0^{(\ell-k)n}) = \pi^{\ell-k} \Big( \pi( \pi(\cdots \pi( \pi(\beta_1) \oplus \beta_2 ) \cdots) \oplus \beta_k ) \Big) = \pi^{\ell-k}( \pi (  \overline{\beta}  ) )  = \pi^{\ell-k+1}(\overline{\beta}) $ until $\pi^{\ell-k  +1 }(\overline{\alpha})$ and $\pi^{\ell-k  +1 }(\overline{\beta})$ are different (which will already be the case after a single try except with negligible probability).
Then, since $\pi^{\ell-k  +1 }(\overline{\alpha}) \neq \pi^{\ell-k  +1 }(\overline{\beta})$, we must also have $\overline{\alpha} \neq \overline{\beta}$.

We have 
\[  \cbc{\ell}{\pi} (\prefisso{\alpha}{k} \concat 0^{(\ell-k)n}) = \cbc{\ell-k+1}{\pi} (\overline{\alpha} \concat 0^{(\ell-k)n})  \]
and similarly
\[  \cbc{\ell}{\pi} (\prefisso{\beta}{k} \concat 0^{(\ell-k)n}) = \cbc{\ell-k+1}{\pi} (\overline{\beta} \concat 0^{(\ell-k)n}) \ . \]

So, for $j=1 \dots \ell-k$ we define 
\begin{align*}
  g_j : \stringhe{n+1} &\to \stringhe{n} \\
  b \concat x \mapsto &\begin{sistema}
    \cbc{\ell}{\pi} (\prefisso{\alpha}{k}\concat 0^{(j-1)n} \concat x \concat 0^{(\ell-k-j)n}) \testo{ if } b=0 \\[1mm]
    \cbc{\ell}{\pi} (\prefisso{\beta}{k} \concat 0^{(j-1)n} \concat x \concat 0^{(\ell-k-j)n}) \testo{ if } b=1
  \end{sistema} \\
   = &\begin{sistema}
    \cbc{\ell-k+1}{\pi} (\overline{\alpha} \concat 0^{(j-1)n} \concat x \concat 0^{(\ell-k-j)n}) \testo{ if } b=0 \\[1mm]
    \cbc{\ell-k+1}{\pi} (\overline{\beta} \concat 0^{(j-1)n} \concat x \concat 0^{(\ell-k-j)n}) \testo{ if } b=1
  \end{sistema} \\ 
  = &\begin{sistema} 
    \pi^{\ell-k+1-j}(\pi^j(\overline{\alpha}) \oplus x) \testo{ if } b=0 \\[1mm] 
    \pi^{\ell-k+1-j}(\pi^j(\overline{\beta}) \oplus x)  \testo{ if } b=1 
  \end{sistema} \, .
\end{align*}
Then we can run the algorithm from Section~\ref{sec:cbc} because
$\ell-k+1\geq 3$. Thereby, we obtain a tag for the message $m=
\prefisso{\beta}{k} \concat z^1 \concat \cdots \concat z^{\ell-k}$,
where $z^j = \pi^j (\overline{\alpha}) \oplus
\pi^j(\overline{\beta})$, and only (superpositions of) messages with at least one zero-block have been queried in the process.

\section{Conclusion and Open Questions}
We have shown two applications of Simon's algorithm to breaking classical symmetric cryptoschemes. As illustrated in the independent work of~\cite{C:KLLN16}, the same attack technique can be applied in various other cryptographic scenarios as well. These results shed new light on the post-quantum cryptographic security of classical cryptography and suggest that many classical security proofs need to be revisited in the light of quantum attackers.

Very recent work by Zhandry~\cite{Zhandry16} shows the existence of quantum-secure pseudo-random permutations (PRPs). However, the construction does not use Feistel networks. Therefore, it remains an intriguing open question to prove that a constant number (strictly larger than 3) of Feistel rounds suffices to obtain a quantum-secure pseudorandom permutation from a quantum secure PRF. In the case of message-authentication codes, our work suggests that one has to be very careful in using popular constructions such as CBC-MAC. 

\nonumsection{Acknowledgements}
\noindent
We would like to thank the authors of~\cite{C:KLLN16} and Tommaso Gagliardoni for useful discussions about this topic, and Ronald de Wolf for feedback on a draft of the article. Large parts of this work was carried out when TS was a master student at ILLC. CS is supported by an NWO VIDI grant.

\nonumsection{References}
\noindent
\bibliographystyle{quantum-bib/bibtex/bst/alphaarxiv}
\bibliography{quantum-bib/bibtex/bib/full,quantum-bib/bibtex/bib/quantum,abbrev2,crypto,quant_ref}

\newcommand{\etalchar}[1]{$^{#1}$}
\begin{thebibliography}{KLLNP16}

\bibitem[BBD09]{BBD09}
D.~J. Bernstein, J.~Buchmann, and E.~Dahmen, editors.
\newblock {\em Post-Quantum Cryptography}.
\newblock Springer, 2009.
\newblock \\
  \texttt{DOI:\,\href{http://dx.doi.org/10.1007/978-3-540-88702-7}{10.1007/978-3-540-88702-7}}.

\bibitem[BDF{\etalchar{+}}11]{BDF+11}
D.~Boneh, {\"{O}}.~Dagdelen, M.~Fischlin, A.~Lehmann, C.~Schaffner, and
  M.~Zhandry.
\newblock Random oracles in a quantum world.
\newblock In {\em Advances in Cryptology---ASIACRYPT 2011}, pages 41--69, 2011.
\newblock \\
  \texttt{DOI:\,\href{http://dx.doi.org/10.1007/978-3-642-25385-0\_3}{10.1007/978-3-642-25385-0\_3}}.

\bibitem[BKR00]{BKR00}
M.~Bellare, J.~Kilian, and P.~Rogaway.
\newblock The security of the cipher block chaining message authentication
  code.
\newblock {\em Journal of Computer and System Sciences}, 61(3): 362--399, 2000.
\newblock \\
  \texttt{DOI:\,\href{http://dx.doi.org/10.1006/jcss.1999.1694}{10.1006/jcss.1999.1694}}.

\bibitem[BZ13a]{EC:BonZha13}
D.~Boneh and M.~Zhandry.
\newblock Quantum-secure message authentication codes.
\newblock In T.~Johansson and P.~Q. Nguyen, editors, {\em EUROCRYPT~2013},
  volume 7881 of {\em {LNCS}}, pages 592--608, Athens, Greece, 2013. Springer,
  Heidelberg, Germany.
\newblock \\
  \texttt{DOI:\,\href{http://dx.doi.org/10.1007/978-3-642-38348-9\_35}{10.1007/978-3-642-38348-9\_35}}.

\bibitem[BZ13b]{BZ13}
D.~Boneh and M.~Zhandry.
\newblock Secure signatures and chosen ciphertext security in a quantum
  computing world.
\newblock In {\em Advances in Cryptology---CRYPTO 2013}, pages 361--379, 2013.
\newblock \\
  \texttt{DOI:\,\href{http://dx.doi.org/10.1007/978-3-642-40084-1\_21}{10.1007/978-3-642-40084-1\_21}}.

\bibitem[DFNS14]{DFNS14}
I.~Damg{\aa}rd, J.~Funder, J.~B. Nielsen, and L.~Salvail.
\newblock Superposition attacks on cryptographic protocols.
\newblock In {\em Information Theoretic Security}, pages 142--161, 2014.
\newblock \\
  \texttt{DOI:\,\href{http://dx.doi.org/10.1007/978-3-319-04268-8\_9}{10.1007/978-3-319-04268-8\_9}}.

\bibitem[DH76]{DH76}
W.~Diffie and M.~Hellman.
\newblock New directions in cryptography.
\newblock {\em IEEE Transactions on Information Theory}, 22(6): 644--654, 1976.
\newblock \\
  \texttt{DOI:\,\href{http://dx.doi.org/10.1109/tit.1976.1055638}{10.1109/tit.1976.1055638}}.

\bibitem[GHS16]{C:GagHulSch16}
T.~Gagliardoni, A.~H{\"u}lsing, and C.~Schaffner.
\newblock Semantic security and indistinguishability in the quantum world.
\newblock In M.~Robshaw and J.~Katz, editors, {\em CRYPTO~2016, Part III},
  volume 9816 of {\em {LNCS}}, pages 60--89, Santa Barbara, CA, USA, 2016.
  Springer, Heidelberg, Germany.
\newblock \\
  \texttt{DOI:\,\href{http://dx.doi.org/10.1007/978-3-662-53015-3\_3}{10.1007/978-3-662-53015-3\_3}}.

\bibitem[Gro96]{STOC:Grover96}
L.~K. Grover.
\newblock A fast quantum mechanical algorithm for database search.
\newblock In {\em 28th ACM STOC}, pages 212--219, Philadephia, PA, USA, 1996.
  {ACM} Press.

\bibitem[KL07]{KL07}
J.~Katz and Y.~Lindell.
\newblock {\em Introduction to Modern Cryptography (Chapman \& Hall/Crc
  Cryptography and Network Security Series)}.
\newblock Chapman \& Hall/CRC, 2007.

\bibitem[KLLNP16]{C:KLLN16}
M.~Kaplan, G.~Leurent, A.~Leverrier, and M.~Naya-Plasencia.
\newblock Breaking symmetric cryptosystems using quantum period finding.
\newblock In M.~Robshaw and J.~Katz, editors, {\em CRYPTO~2016, Part II},
  volume 9815 of {\em {LNCS}}, pages 207--237, Santa Barbara, CA, USA, 2016.
  Springer, Heidelberg, Germany.
\newblock \\
  \texttt{DOI:\,\href{http://dx.doi.org/10.1007/978-3-662-53008-5\_8}{10.1007/978-3-662-53008-5\_8}}.

\bibitem[KM10]{kuwakado2010quantum}
H.~Kuwakado and M.~Morii.
\newblock Quantum distinguisher between the 3-round feistel cipher and the
  random permutation.
\newblock In {\em Information Theory Proceedings (ISIT), 2010 IEEE
  International Symposium on}, pages 2682--2685. IEEE, 2010.

\bibitem[KM12]{kuwakado2012security}
H.~Kuwakado and M.~Morii.
\newblock Security on the quantum-type even-mansour cipher.
\newblock In {\em Information Theory and its Applications (ISITA), 2012
  International Symposium on}, pages 312--316. IEEE, 2012.

\bibitem[LR88]{LubRac88}
M.~Luby and C.~Rackoff.
\newblock How to construct pseudorandom permutations from pseudorandom
  functions.
\newblock {\em {SIAM} Journal on Computing}, 17(2), 1988.

\bibitem[Mer78]{Mer78}
R.~C. Merkle.
\newblock Secure communications over insecure channels.
\newblock {\em Communications of the ACM}, 21(4): 294--299, 1978.
\newblock \\
  \texttt{DOI:\,\href{http://dx.doi.org/10.1145/359460.359473}{10.1145/359460.359473}}.

\bibitem[MW16]{MdW16}
A.~Montanaro and R.~d. Wolf.
\newblock {\em A Survey of Quantum Property Testing}.
\newblock Number~7 in Graduate Surveys. Theory of Computing Library, 2016.
\newblock \\
  \texttt{DOI:\,\href{http://dx.doi.org/10.4086/toc.gs.2016.007}{10.4086/toc.gs.2016.007}}.

\bibitem[RSA78]{RSA78}
R.~L. Rivest, A.~Shamir, and L.~Adleman.
\newblock A method for obtaining digital signatures and public-key
  cryptosystems.
\newblock {\em Communications of the ACM}, 21(2): 120--126, 1978.
\newblock \\
  \texttt{DOI:\,\href{http://dx.doi.org/10.1145/359340.359342}{10.1145/359340.359342}}.

\bibitem[Sho94]{Sho94}
P.~W. Shor.
\newblock Algorithms for quantum computation: discrete logarithms and
  factoring.
\newblock In {\em 35th Annual Symposium on Foundations of Computer
  Science---FOCS 1994}, pages 124--134. IEEE, 1994.
\newblock \\
  \texttt{DOI:\,\href{http://dx.doi.org/10.1109/SFCS.1994.365700}{10.1109/SFCS.1994.365700}}.

\bibitem[Sim94]{FOCS:Simon94}
D.~R. Simon.
\newblock On the power of quantum cryptography.
\newblock In {\em 35th FOCS}, pages 116--123, Santa Fe, New Mexico, 1994.
  {IEEE} Computer Society Press.

\bibitem[War05]{War05}
W.~P. Wardlaw.
\newblock Row rank equals column rank.
\newblock {\em Mathematics Magazine}, 78(4): 316--318, 2005.
\newblock \\ Online: \url{http://www.jstor.org/stable/30044181}.

\bibitem[Zha16]{Zhandry16}
M.~Zhandry.
\newblock A note on quantum-secure prps, 2016.

\end{thebibliography}

% \begin{thebibliography}{000}
% \bibitem{first}
% P. Horodecki and R. Horodecki (2001), {\it Distillation and bound entanglement},
% Quantum Inf. Comput., Vol.1, pp. 045-075.

% \bibitem{cal}
% R. Calderbank and P. Shor (1996), {\it Good quantum error
%        correcting codes exist},
% Phys. Rev. A, 54, pp. 1098-1106.

% \bibitem{niel}
% M.A. Nielsen and J. Kempe (2001), {\it Separable states are
% more disordered globally than locally}, quant-ph/0105090.

% \bibitem{mar}
% A.W. Marshall and I. Olkin (1979), {\it Inequalities: theory of majorization and its applications},
% Academic Press (New York).
% \end{thebibliography}

% \noindent
% Appendices should be used only when absolutely necessary. They
% should come after the References. If there is more than one
% appendix, number them alphabetically. Number displayed equations
% occurring in the Appendix in this way, e.g.~(\ref{that}), (A.2),
% etc.
% \begin{equation}
% \langle\hat{O}\rangle=\int\psi^*(x)O(x)\psi(x)d^3x~. 
% \label{that}
% \end{equation}

\end{document}